\documentclass[journal,twocolumn,10pt,twoside]{IEEEtran}
\normalsize

\ifCLASSINFOpdf
   \usepackage[pdftex]{graphicx}
   \graphicspath{{../pdf/}{../jpeg/}}
   \DeclareGraphicsExtensions{.pdf,.jpeg,.png}
\else
\fi

\usepackage[cmex10]{amsmath}

\usepackage{float}
\usepackage{amsmath}
\usepackage{amssymb}
\usepackage{paracol}
\usepackage{multirow}
\usepackage[nolist,nohyperlinks]{acronym}
\usepackage{bm} 
\usepackage{comment}
\usepackage[table]{xcolor}
\usepackage{array} 
\newcolumntype{L}[1]{>{\raggedright\arraybackslash}m{#1}}
\newcolumntype{C}[1]{>{\centering\arraybackslash}m{#1}}
\newcolumntype{R}[1]{>{\raggedleft\arraybackslash}m{#1}}

\makeatletter
\newcommand{\acx}{\protect\@acx}%
\newcommand{\@acx}[1]{%
  \ifAC@dua
   \acl{#1}%
  \else
   \expandafter\ifx\csname ac@#1\endcsname\AC@used
      \acs{#1}%
   \else
      \acl{#1}%
   \fi
  \fi
}
\makeatother

\makeatletter
\newcommand{\defeq}{\mathrel{\aban@defeq}}
\newcommand{\aban@defeq}{%
	\vbox{\offinterlineskip\check@mathfonts
		\ialign{\hfil##\hfil\cr
			\fontsize{\ssf@size}{\z@}\normalfont def\cr
			\noalign{\kern1\p@}
			$\m@th=$\cr
			\noalign{\kern-.5\fontdimen22\textfont2}
		}%
	}%
}
\makeatother

\begin{acronym}
\acro{SG}{Smart Grid}
\acro{AMR}{automatic meter reading}
\acro{AMI}{advanced metering infrastructure}
\acro{BW}{bandwidth}
\acro{PLC}{power line communications}
\acro{NB-PLC}{narrowband \acx{PLC}}
\acro{OFDM}{orthogonal frequency division multiplexing}
\acro{DFT}{discrete Fourier transform}
\acro{IDFT}{inverse \acx{DFT}}
\acro{ICI}{intercarrier interference}
\acro{ISI}{intersymbol interference}
\acro{EMC}{electromagnetic compatibility}
\acro{AIC}{active interference cancellation}
\acro{CC}{cancellation carriers}
\acro{OOBE}{out-of-band emissions}
\acro{PSD}{power spectral density}
\acro{RC}{raised cosine}
\acro{RHS}{right hand side}
\acro{PAPR}{peak-to-average power ratio}
\acro{w.r.t.}{with respect to}

\acro{LTE}{long-term evolution}
\acro{4G}{fourth generation}
\acro{5G}{fifth generation}
\acro{IoT}{Internet of Things}
\acro{NB-IoT}{narrowband \acx{IoT}}
\acro{CR}{cognitive radio}
\acro{DSL}{digital subscriber line}
\acro{AST}{adaptive symbol transition}
\acro{FIR}{finite impulse response}
\acro{IIR}{infinite impulse response}
\acro{BER}{bit error rate}
\acro{FEQ}{frequency equalizer}
\acro{SVD}{singular value decomposition}
\acro{GSVD}{generalized singular valued decomposition}
\acro{LSNP}{least squares notch precoder}
\acro{LS}{least squares}

\acro{ACLR}{adjacent channel leakage ratio}

\end{acronym}

\DeclareMathOperator*{\argmin}{arg\,min} 

\definecolor{reviewer1}{rgb}{0 0 1}
\definecolor{reviewer2}{rgb}{1 0 0}
\definecolor{reviewer3}{rgb}{0 0.5 0}

\definecolor{general}{rgb}{0.8, 0.33, 0.0}

\begin{document}
%
\title{A modified pulse and design framework to halve the complexity of OFDM spectral shaping techniques}
\title{A Modified Pulse and Design Framework to Halve the Complexity of OFDM Spectral Shaping Techniques}
%
%
%
\author{Javier Giménez, José A. Cortés, Francisco Javier Cañete, Eduardo Martos-Naya and Luis Díez
\thanks{The authors are with the Communications and Signal Processing Lab, Telecommunication Research Institute (TELMA), Universidad de Málaga, E.T.S. Ingeniería de Telecomunicación, Bulevar Louis Pasteur 35, 29010 Málaga, Spain.
Corresponding author: Javier Giménez. E-mail: javierg@ic.uma.es; jaca@ic.uma.es; francis@ic.uma.es; eduardo@ic.uma.es; diez@ic.uma.es. This work has been funded in part by the Spanish Government under project PID2019-109842RB-I00/AEI/10.13039/501100011033 and FPU grant FPU20/03782, by the European Fund for Regional Development (FEDER), Junta de Andalucía and the Universidad de Málaga under projects P18-TP-3587 and UMA20-FEDERJA-002.}}

\maketitle

\begin{abstract}
\Ac*{OFDM} is a widespread modulation but suffers from high \ac*{OOBE}. Spectral shaping strategies such as precoding, \ac*{AIC} and time-domain methods are effective at reducing the OOBE but entail optimization procedures and real-time implementation costs which might be considerable. This paper proposes a modification of the conventional OFDM waveform aimed at reducing the cost associated to many of the state-of-the-art spectral shaping techniques and sets a framework for future works that want to benefit from the same reduction. This approach may reduce both the number of coefficients involved in the optimization and the number of products of its implementation by up to 50\%.
\end{abstract}

\vspace{-0.2cm}

%
\IEEEpeerreviewmaketitle

\section{Introduction}
\label{intro}
\IEEEPARstart{T}{he} need for an efficient use of the spectrum is forcing increasingly stricter \ac{OOBE} masks that allow reducing the guard bands and a dynamic sharing of the spectral resources \cite{Zaidi2016}\cite{Haykin05}. While \ac{OFDM} is broadly used in wireless and wired systems, it shows poor spectral confinement. This can be mitigated by employing wide guard bands and by nulling the data carriers located by the edges of the notched bands, to the detriment of the system throughput. To avoid this, several spectral shaping methods have been proposed. Most of them can be classified into two groups according to the domain in which they are applied: frequency-domain and time-domain. Details of the main techniques in each category can be found in \cite{You14}\cite{Hussain2022} and references therein. The solutions yielded by these techniques require costly optimization procedures and high real-time implementation costs, which are sometimes alleviated by means of method-specific strategies. 

Motivated by the lack of complexity reduction techniques of general validity that can be applied to all types of \ac{OFDM} spectral shaping methods, this work proposes a modification of the conventional \ac{OFDM} signal that allows diminishing the computational cost of state-of-the-art techniques by up to 50\%. It also defines a design framework that guarantees that future techniques designed according to it can benefit from this reduction. In summary, the contributions of this paper are:
\begin{itemize}
    \item The proposal of a modified pulse for \ac{OFDM} systems that allows reducing the complexity of existing spectral shaping techniques originally designed for  conventional \ac{OFDM} signals. The new pulse entails marginal modifications to the transmiter. Moreover, the resulting waveform can be transparent to the receiver, as the change relative to the conventional one can be absorbed by the \ac{FEQ} typically used in \ac{OFDM} receivers. 

    \item The definition of a design framework for spectral shaping techniques consisting in a cost function and a set of possible constraints. Designing according to this framework and using the proposed pulse is a sufficient condition for new techniques to take advantage of a notable complexity reduction in the optimization and real-time implementation processes. However, some existing methods that do not fit in the framework still might benefit from this reduction just by modifying the proposed pulse.

    \item It provides a compilation of existing precoding, \ac{AIC} and \ac{AST} spectral shaping methods that can exploit our proposals and quantifies the attainable complexity reduction, which can be up to 50\%.
\end{itemize}

\section{Background: The Power Spectral Density of OFDM Signals}
\label{sec1}
The discrete-time low-pass equivalent expression of an \ac{OFDM} signal is
\begin{equation}
x(n) = \sum_{u=-\infty}^{\infty} x_u(n-uN_{\textrm{s}}),
    \label{eq:sec1:senalOFDMLPeq}
\end{equation}
where $N_{\textrm{s}} = N+N_{\textrm{GI}}$ is the symbol period, $N$ is the size of the \ac{IDFT} and $N_{\textrm{GI}}$ is the length of the guard interval. The $u$-th symbol is given by
\begin{equation}
x_u(n) = \sum_{k\in \mathcal{D}} p_k(n) d_k(u)
    \label{eq:sec1:simbOFDM}
\end{equation}
where $\mathcal{D}=\left\{ d_1, \dots, d_{|\mathcal{D}|} \right\}$ denotes the set of carriers that convey data in the \ac{OFDM} system (data carriers) and $d_k(u)$ and $p_k(n)$ are the $u$-th modulating symbol and the base pulse employed in carrier $k$, respectively, with
\begin{equation}
p_k(n) = w(n) e^{j\frac{2\pi}{N}k (n-N_{\textrm{GI}})},   \quad k\in \mathcal{D}.
    \label{eq:sec1:pulsoBase}
\end{equation}
For generality, the pulse $p_k(n)$ is shaped by the discrete-time window function $w(n)$, which is non-zero only in the interval $n\in \left\{ 0, \dots, L-1 \right\}$ with $L=N_{\textrm{s}}+\beta$ and can have smooth transitions of $\beta$ samples at both ends. The \ac{RC} transition is typically used to this end. Note that using this type of pulses reduces the effective length of the cyclic prefix by $\beta$ samples due to the tapered transitions at the beginning of the pulse and the overlap with the preceding \ac{OFDM} symbol.


Considering that the symbols $d_k(u)$ are white for a given $k$ and independent for different $k$, an analytical expression for the \ac{PSD} of (\ref{eq:sec1:senalOFDMLPeq}) can be obtained, 
\begin{equation}
S(f) = \frac{1}{N_{\textrm{s}}} \sum_{k\in \mathcal{D}} \sigma_k^2 |P_k(f)|^2,
    \label{eq:sec1:PSDsenalConvencional}
\end{equation}
where $\sigma_k^2$ is the variance of the sequence $d_k(u)$, and $P_k(f)$ is the Fourier transform of $p_k(n)$. The latter can be expressed as 
\begin{equation}
P_k(f) = \mathbf{f}_L^{\textrm{H}}(f) \mathbf{p}_k,
    \label{eq:sec1:TFpulsoBase}
\end{equation}
where $\mathbf{p}_k$ is vector form of the basic pulse with $L \times 1$ samples, $\mathbf{f}_L(f) = \left[ 1, e^{j2\pi f}, \dots, e^{j2\pi f(L-1)} \right]^{\textrm{T}}$ and $f \in [-1/2, 1/2)$.

\section{Spectral shaping techniques}
The signals generated by the most common spectral shaping methods can be expressed as particular cases of the general formulation obtained by replacing $p_k(n)$ by the  pulse
\begin{equation}
h_k(n) = p_k(n) + \!\!\!\sum_{i \in \mathcal{K}; i \neq k}{\!\!\!g_{i,k} p_i(n)} + t_k(n), \quad k\in \mathcal{D},
    \label{eq:sec1:pulsoGeneral}
\end{equation}
where the set $\mathcal{K}$ contains the indexes of the active carriers, i.e., with allocated power. In most spectral shaping methods, $\mathcal{K}$ is the union of the set of data carriers, $\mathcal{D}$, and the set of redundant carriers, $\mathcal{C}=\left\{ c_1, \dots, c_{|\mathcal{C}|} \right\}$. While carriers in $\mathcal{D}$ may be exclusively used to convey data or may have the dual functionality of conveying information and shaping the spectrum, carriers in $\mathcal{C}$ are only used to achieve better spectral confinement or to improve the receiver operation at the expense of a lower spectral efficiency. To avoid the latter, some methods do not employ redundant carriers \cite{vandeBeek09}.

The first term on the \ac{RHS} of (\ref{eq:sec1:pulsoGeneral}) is the base pulse, followed by two terms aimed at reducing the \ac{OOBE} produced by the former. The second term achieves this end by means of a weighted sum, with $g_{i,k} \in \mathbb{C}$, of the pulses transmitted in the remaining carriers. Finally, the last term in (\ref{eq:sec1:pulsoGeneral}), $t_k(n)$, consists in a discrete-time sequence that is non-zero only in the range $n\in \left\{ 0, \dots, \beta-1, L-\beta, \dots, L-1 \right\}$, overlapping with the first and last $\beta$ samples of the pulse. This term reduces the \ac{OOBE} by smoothing the symbol boundaries.

The following well-known \ac{OOBE} reduction strategies can be obtained by particularizing the pulse in (\ref{eq:sec1:pulsoGeneral}) as follows:

\subsubsection{\Acf{AIC}}
In these techniques, the redundant carriers in $\mathcal{C}$ are referred to as \ac{CC}, and are used to reduce the \ac{OOBE} produced by the set of carriers that are exclusively used to convey data, $\mathcal{D}$. The resulting pulse is obtained by particularizing (\ref{eq:sec1:pulsoGeneral}) as
\begin{equation}
h_k(n) = p_k(n) + \sum_{i\in \mathcal{C}}g_{i,k} p_i(n), \quad k \in \mathcal{D},
    \label{eq:sec1:pulsoAIC}
\end{equation}
where the term $t_k(n)$ has been left out.

Note that $\mathcal{K} = \mathcal{D} \cup \mathcal{C}$ and $\mathcal{D}\cap \mathcal{C} = \emptyset$, so data carriers are exclusively employed to convey data and cancellation carriers are used to suppress the \ac{OOBE} produced by the former.

\subsubsection{Precoding}
This spectral shaping strategy generalizes \ac{AIC} techniques by extending the set of carriers used to reduce the \ac{OOBE} to the whole set of active carriers, $\mathcal{K}$. The employed pulse can be generically written as
\begin{equation}
h_k(n) = \sum_{i \in \mathcal{K}} g_{i,k} p_i(n), \quad k\in \mathcal{D},
    \label{eq:sec1:pulsoPrecoding}
\end{equation}
where commonly $g_{k,k}=1$. In these circumstances, (\ref{eq:sec1:pulsoPrecoding}) can be obtained from (\ref{eq:sec1:pulsoGeneral}) by grouping the two first terms on the \ac{RHS} of (\ref{eq:sec1:pulsoGeneral}) and leaving out $t_k(n)$. In matrix form, the pulse $\mathbf{h}_k = \left[ h_k(0), \dots, h_k(L-1) \right]$ is expressed as $\mathbf{h}_k =  \mathbf{P}_{\mathcal{K}} \mathbf{g}_k$, for $k \in \mathcal{D}$,
where $\mathbf{g}_{k} = \left[ g_{i_0,k}, \dots, g_{i_{|\mathcal{K}|-1},k} \right]^{\textrm{T}}$ and $\mathbf{P}_{\mathcal{K}} = \left[ \mathbf{p}_{i_0}, \dots, \mathbf{p}_{i_{|\mathcal{K}|-1}} \right]$. Accordingly, the $u$-th \ac{OFDM} symbol in (\ref{eq:sec1:simbOFDM}) can be expressed in matrix form as 
\begin{equation}
\mathbf{x}_u = \mathbf{H}_{\mathcal{D}} \mathbf{d}(u) = \mathbf{P}_{\mathcal{K}} \mathbf{G} \mathbf{d}(u),
    \label{eq:sec1:simbOFDMprecoding}
\end{equation}
where $\mathbf{H}_{\mathcal{D}} = \left[ \mathbf{h}_{k_0}, \dots, \mathbf{h}_{k_{|\mathcal{D}|-1}} \right]$, $\mathbf{d}(u) = \left[ d_{k_{0}}(u), \dots, d_{k_{|\mathcal{D}|-1}}(u) \right]^{\textrm{T}}$ and $\mathbf{G} = \left[ \mathbf{g}_{k_0}, \dots, \mathbf{g}_{k_{|\mathcal{D}|-1}} \right]$, with (possibly) $ \mathbf{G}\left[k,k \right] = 1, \; \forall k \in \mathcal{D}$.

\subsubsection{\Acf{AST}}
The \ac{OOBE} is reduced by adaptively designing the transitions of the \ac{OFDM} symbols. They can be obtained by particularizing (\ref{eq:sec1:pulsoGeneral}) as
\begin{equation}
h_k(n) = p_k(n) + t_k(n), \quad k\in \mathcal{D}.
    \label{eq:sec1:pulsoAST}
\end{equation}

The term $t_k(n)$, herein referred to as \emph{transition pulse}, overlaps with the cyclic prefix of the current and next \ac{OFDM} symbol, which is discarded by the receiver, so it does not interfere with the regular reception of the symbols. As seen, every data carrier has its own embedded transition pulse.

The coefficients of the cancellation terms in all the presented strategies are obtained through an optimization procedure, whose cost function and  constraints differ in the numerous works that can be found in the literature.
\section{Modified OFDM pulse and proposed optimization framework}
\label{sec2}
We propose the following novel pulse to replace the conventional one, $p_k(n)$, used to generate the \ac{OFDM} symbols,
\begin{equation}
\widetilde{p}_k(n) = w(n+\eta) e^{j\frac{2\pi}{N}kn},
    \label{eq:sec2:pulsoBaseHerm}
\end{equation}
where $\eta = (L-1)/2$. Since $w(n+\eta)$ has Hermitian symmetry, so does $\widetilde{p}_k(n)$. This pulse differs from the one used in actual \ac{OFDM} systems, given in (\ref{eq:sec1:pulsoBase}), in a time advance by $\eta$, which is irrelevant since the time origin is arbitrarily fixed, and in that the phase origin for every carrier is now at the central sample. This simple modification of the transmitted signal makes the Fourier transform of $\widetilde{p}_k(n)$, denoted by $\widetilde{P}_k(f)$, to be real valued. This has a direct impact on the optimization and real-time implementation complexity of the spectral shaping method solutions, as it will be demonstrated in section \ref{sec3}. 

This new Hermitian-symmetric pulse can be used to replace the conventional $p_k(n)$ in (\ref{eq:sec1:pulsoGeneral}), leading to,
\begin{equation}
\widetilde{h}_k(n) = \widetilde{p}_k(n) + \!\!\!\sum_{i\in \mathcal{K}; i \neq k}\!\!\! \widetilde{g}_{i,k} \widetilde{p}_i(n) + \widetilde{t}_k(n), \quad k \in \mathcal{D}.
    \label{eq:sec2:pulsoGeneralizadoSim}
\end{equation}

The \ac{PSD} of the resulting \ac{OFDM} signal when the data carriers in $\mathcal{D}$ employ the pulse $\widetilde{h}_k(n)$ can be expressed as 
\begin{equation}
S(f) = \frac{1}{N_{\textrm{s}}} \sum_{k\in \mathcal{D}} \sigma_k^2 |\widetilde{H}_k(f)|^2,
    \label{eq:sec2:PSDPulsoGeneral}
\end{equation}
where $\widetilde{H}_k(f)$ is the Fourier transform of $\widetilde{h}_k(n)$. 

Let's consider the mask function $M_0(f)$, such that $M_0(f)=1$ for the frequencies where the \ac{OOBE} needs to be reduced, i.e., the frequencies in the notched band, and $M_0(f)=0$ for those inside the passband. However, values in between $0$ and $1$ can be used to custom weigh the \ac{OOBE} reduction goal in different subbands separately. The \ac{OOBE} produced by the \ac{OFDM} signal is then given by
\begin{equation}
\begin{aligned}
    P_{\textrm{OOBE}} &= \!\! \int_{-1/2}^{1/2} \!\!\!\!\!\!\!\! M_0(f) S(f) df = \frac{1}{N_{\textrm{s}}} \! \sum_{k \in \mathcal{D}} \sigma_k^{2} \int_{-1/2}^{1/2} \!\!\!\!\!\!\!\! M_0(f) \left|\widetilde{H}_k(f)\right|^2 \!\! df.
\end{aligned}
    \label{eq:sec2:expresionOOBE}
\end{equation}

Now, we propose a general optimization problem that, along with the pulse in (\ref{eq:sec2:pulsoGeneralizadoSim}), configures a spectral shaping framework that allows reducing the complexity of both the optimization and the real-time implementation cost. To this end, the matrix $\widetilde{\mathbf{H}}_{\mathcal{D}} = \left[ \widetilde{\mathbf{h}}_{k_0}, \dots, \widetilde{\mathbf{h}}_{k_{|\mathcal{D}|-1}} \right]$ is defined, which contains the set of discrete-time pulses used by the data carriers arranged column-wise. The optimization problem is then stated as
\begin{equation}
\begin{aligned}
    &\widehat{\mathbf{H}}_{\mathcal{D}} = \argmin_{\widetilde{\mathbf{H}}_{\mathcal{D}}} \left\{ \sum_{k \in \mathcal{D}} \sigma_k^2 \int_{-1/2}^{1/2} M_0(f) \left| \widetilde{H}_k(f) \right|^2 df\right\}\\
    &\textrm{s.to.}\\
    &\;\;\;\; \sum_{k \in \mathcal{D}} \sigma_k^2 \int_{-1/2}^{1/2} \!\!\!\!\!\! M_j(f) \left| \widetilde{H}_k(f) \right|^2 df \leq \delta_j, \quad 1 \leq j\leq C
\end{aligned}
    \label{eq:sec2:problemaOpt}    
\end{equation}
where $M_j(f)$ is the $j$-th mask function, $\delta_j$ the value to which the integral is bounded, and $C$ is the number of constraints that will be applied. Typically, $M_j(f)$ will be defined such that they equal $1$ in the frequency subband where the energy emissions have to be controlled and $0$ elsewhere. Nonetheless, values in between can be considered, as it was previously stated.

Note that both the cost function and the constraints refer to the power of the \ac{OFDM} signal that is emitted at the frequencies indicated by the mask functions $M_j(f)$ when the pulse in (\ref{eq:sec2:pulsoGeneralizadoSim}) is used. In practice, the goal is to minimize the power emitted in the notched band while the constraints are set to, for instance, limit the total transmission power of the signal (when $M_1(f)=1$ $\forall f \in [-1/2, 1/2)$), or avoid \ac{PSD} peaks in the passband, a recurring restriction when using \ac{AIC} techniques \cite{Hussain2022}\cite{Diez19}, among others.

It will be now shown that spectral shaping techniques in which the pulse in (\ref{eq:sec2:pulsoGeneralizadoSim}) is optimized according to (\ref{eq:sec2:problemaOpt}) yield real-valued $\widetilde{g}_{i,k}$. To this end, $\left| \widetilde{H}_k(f) \right|^2$ is expanded as
\begin{equation}
\left| \widetilde{H}_k(f) \right|^2 = \left( \widetilde{H}_k^{\Re}(f) \right)^2 + \left( \widetilde{H}_k^{\Im}(f) \right)^2,
    \label{eq:sec2:moduloPulsoHk}
\end{equation}
where, given that $\widetilde{P}_k(f)$ is real-valued,
\begin{equation}
\begin{aligned}
    &\widetilde{H}_k^{\Re}(f) = \widetilde{P}_k(f) + \sum_{i\in \mathcal{K}; i\neq k} \widetilde{g}_{i,k}^{\Re} \widetilde{P}_i(f) + \widetilde{T}_k^{\Re}(f),\\
    &\widetilde{H}_k^{\Im}(f) =  \sum_{i\in \mathcal{K}; i\neq k} \widetilde{g}_{i,k}^{\Im} \widetilde{P}_i(f) + \widetilde{T}_k^{\Im}(f).
\end{aligned}
    \label{eq:sec2:partesRealImagPulsoHk}
\end{equation}

Minimizing the value of (\ref{eq:sec2:moduloPulsoHk}) in the notched band minimizes the \ac{OOBE} in (\ref{eq:sec2:problemaOpt}). To achieve the former, $\widetilde{H}^{\Im}_k(f)$ must be zero, which can be accomplished by forcing $\widetilde{g}_{i,k}$ and $\widetilde{T}_k(f)$ to be real-valued, with $i\in \mathcal{K},\; k\in \mathcal{D}$. By means of the Fourier transform properties, $\widetilde{T}_k(f)$ being real-valued leads to $\widetilde{t}_k(n)$ having Hermitian symmetry. It can be proven that the  constraints in (\ref{eq:sec2:problemaOpt}) do not interfere with this conclusion, as they also benefit from disregarding $\widetilde{H}_k^{\Im}(f)$.

The previous demonstration can be summarized as follows: when the pulse in (\ref{eq:sec2:pulsoGeneralizadoSim}) is used and an optimization problem like the one in (\ref{eq:sec2:problemaOpt}) is considered, the coefficients $\widetilde{g}_{i,k}$ are real-valued and the term $\widetilde{t}_k(n)$ has Hermitian symmetry. This considerably reduces the computational cost of the optimization procedure, as the number of variables is halved. The real-time implementation complexity is also notably diminished as well, reducing the number of required real products by a half with respect to the complex and non-symmetric case.

It must be noted that complying with the optimization problem in (\ref{eq:sec2:problemaOpt}) is a sufficient but not a necessary condition for the complexity reduction to apply. Hence, proposals that do not fit in (\ref{eq:sec2:problemaOpt}) still can exploit this result when a Hermitian-symmetric pulse like the one in (\ref{eq:sec2:pulsoBaseHerm}) is employed. In particular, many precoding-based methods which impose constraints on the optimization problem aimed at preserving the receiver's correct operation can still take advantage of this reduction. Existing methods that fulfill the latter are shown in Section \ref{sec3}, along with others that also enjoy this perk despite being non-compliant with the problem in (\ref{eq:sec2:problemaOpt}).

\section{Complexity Reduction Summary}
\subsection{Practical aspects of the transmitter implementation}
Using the pulse in (\ref{eq:sec2:pulsoBaseHerm}) requires minor changes in the \ac{OFDM} transmitter. They are aimed at placing the phase origin of all carriers at the central sample, since the time advance is not a problem. It can be easily implemented by applying a circular shift to the right on the samples obtained from the \ac{IDFT}. This shift, of magnitude $(N-N_{\textrm{GI}}+\beta-1)/2$, reorders the samples before the cyclic prefix is added without additional computational cost at the transmitter. This simple implementation is only possible if the extension of the discrete-time pulse, $L$, is an odd number. Otherwise, the phase origin of the Hermitian-symmetric pulse would sit between samples, hindering the possibility of achieving symmetry by reordering the \ac{IDFT}'s output samples. Nevertheless, the cases in which $L$ is an even number can be easily resolved by extending the pulse in just one sample, forcing $\beta$ to be odd. This has almost no impact on the effectiveness of the cyclic prefix, which is reduced in one additional sample.

Reordering the samples of the \ac{OFDM} symbol provokes a rotation of the data carriers' constellations at the receiving end. Strictly compensating this rotation (e.g., performing on the samples of the received signal a circular shift to the left of the same magnitude as in the transmitter) requires the receiver to be aware of the spectral shaping. However, since this rotation could be also estimated as part of the channel response and absorbed by the \ac{FEQ} typically employed in \ac{OFDM} systems, in practice, this modification might be transparent to the receiver.

Finally, it must be highlighted that \ac{OFDM} systems that do not use the Hermitian-symmetric waveform in (\ref{eq:sec2:pulsoGeneralizadoSim}), but in which the optimization can be expressed as in (\ref{eq:sec2:problemaOpt}), can still benefit from the presented results and get a notable reduction in the complexity of the optimization process (yet not in the real-time implementation one). The reason is that the solution of (\ref{eq:sec2:problemaOpt}) with the non-symmetric pulse in (\ref{eq:sec1:pulsoGeneral}) can be computed from the solution obtained with (\ref{eq:sec1:pulsoGeneral})  as
\begin{equation}
\begin{aligned}
    &g_{i,k} = \widetilde{g}_{i,k} e^{j\frac{2\pi}{N}(k-i)(\eta-N_{\textrm{GI}})}\quad\quad\quad
    \mathbf{t}_k = \widetilde{\mathbf{t}}_k e^{j\frac{2\pi}{N}k (\eta-N_{\textrm{GI}})}
\end{aligned}
    \label{eq:sec3:transformSoluciones}
\end{equation}
where $\mathbf{t}_k$ and $\widetilde{\mathbf{t}}_k$  contain the samples of $t_k(n)$ and $\widetilde{t}_k(n)$ in matrix form, respectively. Proof of (\ref{eq:sec3:transformSoluciones}) can be easily obtained.

\subsection{Real-time implementation complexity reduction}
\label{sec4}
This subsection details the complexity reduction obtained when precoding, \ac{AIC} and \ac{AST} methods use the pulse and the optimization framework proposed in this work. The computational cost is given in terms of the number of real products per \ac{OFDM} symbol required to implement the considered spectral shaping method. The key element is that, a product with complex operands requires 4 real products, but only 2 real products when one operand is real-valued. Likewise, if a complex value is to be multiplied by a complex conjugate pair in two separate operations, just 4 real products are required. This reduces the real-time implementation to:

\subsubsection{AIC techniques}
In \cite{Diez19} we designed a pulse that generalized other \ac{AIC} proposals. The implementation complexity analysis yields that the number of additional real products is reduced from $4|\mathcal{D}||\mathcal{C}|$ \cite[table I]{Diez19} to $2|\mathcal{D}||\mathcal{C}|$, a $50$\% reduction.

\subsubsection{Precoding}
It can be deduced from (\ref{eq:sec1:simbOFDMprecoding}) that the amount of additional real products required to implement the spectral precoder is $4|\mathcal{K}||\mathcal{D}|$, as a result of multiplying the $|\mathcal{K}| \times |\mathcal{D}|$ precoding matrix, $\mathbf{G}$, and the $|\mathcal{D}|\times 1$ vector of data $\mathbf{d}(u)$. If matrix $\mathbf{G}$ is real-valued, this number is reduced by $50$\%.

\subsubsection{AST techniques}
The pulse in \cite{Diez19} generalizes many \ac{AST} techniques, for which three implementation alternatives are given, two of them aimed at reducing the complexity. The latter can benefit from the proposal in this work as follows:
\begin{itemize}
    \item Regular transition pulse: the number of real products is reduced from $8\beta |\mathcal{D}|$ \cite[table I]{Diez19} to $4\beta |\mathcal{D}|$, i.e., a $50$\% reduction is achieved.
    \item Harmonically designed transition pulse: the complexity reduction depends on the number of harmonics used to compose the transition term, denoted as $b$. The number of real products is reduced from $8|\mathcal{D}|b + 2\beta \textrm{log}_2(\beta)$ to $4|\mathcal{D}|b + 2 \beta \textrm{log}_2(\beta)$, which is lower than $50$\% but still relevant.
\end{itemize}

\section{Application to Existing Spectral Shaping Methods}
\label{sec3}
A great number of the \ac{OOBE} methods in the literature can benefit from using $\widetilde{p}_k(n)$ in (\ref{eq:sec2:pulsoBaseHerm}). A selection is listed in Table \ref{table:sec3:tablaMetodos}. Some of them approach the optimization problem in a very similar way to the one in (\ref{eq:sec2:problemaOpt}). However, since the latter is a sufficient condition, other methods listed in Table \ref{table:sec3:tablaMetodos} do not comply with the proposed optimization framework but still enjoy the complexity reduction as long as the Hermitian-symmetric waveform in (\ref{eq:sec2:pulsoBaseHerm}) is employed.

\begin{table*}[t]\centering
  \caption{Spectral shaping methods for \ac{OFDM} signals that would benefit from using the Hermitian-symmetric pulse proposed in this work.}
  \begin{tabular}{|L{2.5cm}|c|}
  \hline
    \bf{Reference(s)}  & \parbox[c][0.5cm]{14.65cm}{\centering \bf{Proof}}\\ 
    \hline
     \parbox[c]{2.4cm}{\vspace{-7pt}\cite[eqs. (18)-(20)]{Diez19}\\ \cite[eqs. (6)-(8)]{Yamaguchi04}\\ \cite[eq. (3)]{Brandes06}\\ \cite[eqs. (12)-(13)]{Gimenez2023}} & \parbox[c][1.4cm]{14.65cm}{The optimization in \cite[eqs. (18)-(19)]{Diez19} and \cite[eqs. (12)-(13) and (30)-(31)]{Gimenez2023} are essentially identical to the one in (\ref{eq:sec2:problemaOpt}). Hence, when $\widetilde{p}_k(n)$ in (\ref{eq:sec2:pulsoBaseHerm}) is used, the optimal coefficients of the \ac{AIC} terms must be real-valued and the optimal \ac{AST} term must have Hermitian-symmetry. Since \cite{Diez19} generalizes \cite{Yamaguchi04} and \cite{Brandes06}, as proven therein, the latter also benefit from this perk.}\\ 
    \hline
    \parbox[c]{2.1cm}{\vspace{-5pt}\cite[eq. (9)]{Hussain2019}\\ \cite[eq. (10)]{Hussain2019a}\\ \cite[eq. (27)]{Hussain2024}} & \parbox[c][1.1cm]{14.65cm}{Their optimization problems are very similar to the one in (\ref{eq:sec2:problemaOpt}), plus the constraints imposed to preserve the correct operation of the receiver. The problems are solved by means of iterative algorithms in which, given that all the terms are real-valued (out-of-band radiation matrices; identity, selection and permutation matrices; etc.), lead to real-valued solutions. }\\
    \hline
    \cite[eqs. (17)-(18)]{Hussain2022} & \parbox[c][1.6cm]{14.65cm}{It proposes two spectral shaping problems: \cite[eq. (17)]{Hussain2022} for orthogonal precoding and \cite[eq. (18)]{Hussain2022} for an \ac{AIC}-based solution. For the first one, the optimal precoding matrix is real-valued when $\widetilde{p}_k(n)$ in (\ref{eq:sec2:pulsoBaseHerm}) is used because it is made of the eigenvectors of a matrix that will be real and symmetric. For the second, the optimal matrix is also computed out of the same real-valued matrix, an identity matrix and a selection matrix, which are also real-valued. In both cases, a shaping window that can be proven to have Hermitian symmetry when using $\widetilde{p}_k(n)$ in (\ref{eq:sec2:pulsoBaseHerm}) is employed.}\\
    \hline
    \cite[eq. (16)]{Kumar2021} & \parbox[c][0.8cm]{14.65cm}{The optimal precoding matrix is made up of the eigenvectors of the matrix in \cite[eq. (16)]{Kumar2021}, which can be easily proved to be real-valued and symmetric, as it results from adding matrices that are real and symmetric when $\widetilde{p}_k(n)$ in (\ref{eq:sec2:pulsoBaseHerm}) is employed.}\\
    \hline
    \cite[eq. (12)]{Kumar2016} & \parbox[c][0.7cm]{14.65cm}{The closed-form expression for the optimal precoding matrix is obtained by inverting a matrix that can be easily proved to be real-valued, as it results from the addition of matrices computed out of the Fourier transform of $\widetilde{p}_k(n)$ in (\ref{eq:sec2:pulsoBaseHerm}).}\\
    \hline
    \cite[eq. (15)]{Zhou2013} & \parbox[c][0.7cm]{14.65cm}{The closed-form expression of the optimal precoding matrix per user consists of a subset of the right-singular vectors of a real-valued matrix. Hence, the optimal matrix is real-valued.}\\
    \hline
    \parbox[c]{2.1cm}{\cite[below eq. (7)]{Kumar2015}\\ \cite[eq. (12)]{vandeBeek09}}  & \parbox[c][1.1cm]{14.65cm}{The optimal precoding matrix in \cite[below eq. (7)]{Kumar2015} is real-valued matrix because  it is obtained as the product of real-valued matrices (an identity matrix, a matrix of real weights and one that contains the Fourier transform of the pulses). In \cite[eq. (12)]{vandeBeek09} only an identity matrix and the one containing the Fourier transform of the pulses appear, and both are real-valued.}\\
    \hline
    \cite[eq. (12)]{Ma2011} & \parbox[c][0.7cm]{14.65cm}{The optimal precoding matrix is real-valued, as it is obtained out of the \ac{SVD} of a matrix that contains the Fourier transform of $\widetilde{p}_k(n)$ in (\ref{eq:sec2:pulsoBaseHerm}), and an arbitrary matrix, which can be chosen to be real as well.}\\
    \hline
  \end{tabular}
  \label{table:sec3:tablaMetodos}
\end{table*}

We now validate our proposal by computing the \ac{PSD} of several methods from Table \ref{table:sec3:tablaMetodos} in two cases: using regular \ac{OFDM} pulses and $\widetilde{p}_k(n)$ in (\ref{eq:sec2:pulsoBaseHerm}). We have selected the methods that combine \ac{AIC} and \ac{AST} proposed by Díez {\emph et al.} \cite{Diez19} and by Hussain and López-Valcarce \cite{Hussain2022}; the spectral precoding techniques by Zhou \emph{et al.} \cite{Zhou2013} (particularized to the single-user case), by van de Beek \cite{vandeBeek09} and by Ma \emph{et al.} \cite{Ma2011}. Interestingly, \cite{Zhou2013} and \cite{vandeBeek09} yield different precoding matrices but lead to the same \ac{PSD}, although different \ac{BER} \cite{Zhou2013}. It must be emphasized that the aim of this assessment is not to compare the performance of the referred techniques, but to prove that their \ac{PSD} remains unaltered when using (\ref{eq:sec2:pulsoBaseHerm}).

The \ac{PSD} of the different methods is obtained using the analytical expressions in \cite[eq. (8)]{Diez19}, \cite[eq. (10)]{Hussain2022}, \cite[eq. (10)]{Zhou2013} and \cite[eq. (6)]{vandeBeek09}, respectively. While \cite{Ma2011} does not provide an analytical expression for the \ac{PSD}, it can be obtained using \cite[eq. (10)]{Zhou2013}. The considered system uses $N = 4096$, $N_{\textrm{GI}} = 1024$.  The methods that use \ac{AST} employ $\beta = 511$. A scenario where the \ac{OOBE} is to be reduced in the bands corresponding to carrier indexes $\mathcal{B} = \left\{ 0, \dots, 1024 \right\} \cup \left\{ 3022, \dots, 3026 \right\} \cup \left\{ 3072, \dots, 4095 \right\}$ is assumed \cite{Diez19}. 

The method in \cite{Diez19} uses a shaping pulse $w(n)$ with \ac{RC} transitions, $3$ \ac{CC} by each edge of the passband ($2$ inband and $1$ out-of-band) and regular transition pulses. This same set of \ac{CC} is used with \cite{Hussain2022}, whose solution is obtained after $16$ iterations with the regularization term configured to avoid \ac{PSD} peaks in the passband. The techniques in \cite{Zhou2013} and \cite{Ma2011} are configured with a coding rate of $\lambda = \frac{2026}{2042}$. The former uses the following set of $16$ normalized frequencies to notch the \ac{PSD}, $\phi = \{ -\phi_1, 0.2378, 0.2379, 0.2380, 0.2385, 0.2386, 0.2387, \phi_1\}$, with $\phi_1=\{0.250, 0.2501, 0.2502, 0.2515, 0.2518\}$, which is also used for \cite{vandeBeek09}. For the method in \cite{Ma2011}, the notched band $\mathcal{B}$ is evenly sampled with $10$ samples per subcarrier spacing.

\begin{figure}[!ht]
    \centering
    \includegraphics[width=\columnwidth]{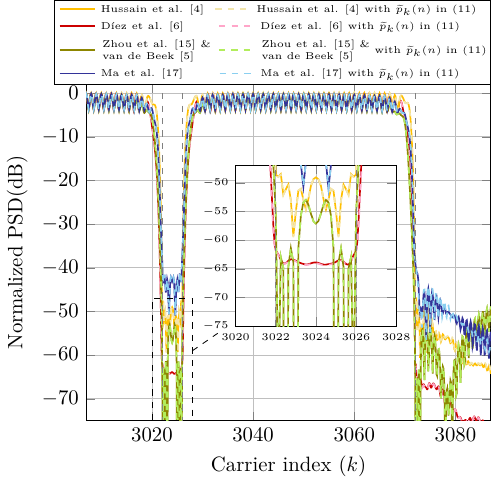}
    \vspace{-0.7cm}
    \caption{Normalized \ac{PSD} given by several methods taken from Table \ref{table:sec3:tablaMetodos} when using the conventional \ac{OFDM} pulse and $\widetilde{p}_k(n)$ in (\ref{eq:sec2:pulsoBaseHerm}). Since \cite{Zhou2013} and \cite{vandeBeek09} yield the same \ac{PSD}, a single curve is depicted for both of them.}
    \vspace{-0.4cm}
    \label{fig:figMascaraTCOM}
\end{figure}

Results are displayed in Fig. \ref{fig:figMascaraTCOM}, which shows that the \ac{PSD} of the assessed methods is unaffected when using the proposed pulse. Similarly, their \ac{PAPR} remains unaltered, although results are not shown for conciseness. On the other hand, employing the proposed pulse allows implementing the 5 considered spectral shaping techniques with just $50\%$ of the number of real products per symbol needed with the conventional \ac{OFDM} pulse.


\section{Conclusion}
\label{conc}
This article has presented a novel pulse and a framework for the design of the spectral shaping methods most typically employed by \ac{OFDM} systems (precoding, \ac{AIC} and \ac{AST}). By exploiting the symmetry of the \ac{OFDM} signal, it halves the number of variables of the optimization procedure and reduces the real-time implementation cost by up to $50$\%. The proposal could be also of interest for the design of future \ac{OFDM}-based communication systems, as it can be easily implemented by adding a circular shift to the right/left in the transmitter/receiver of the conventional \ac{OFDM} system and, in return, the \ac{OOBE} can be reduced with much lower computational cost.

\vspace{-0.1cm}
\section*{Acknowledgment}
The authors thank Maxlinear Hispania S.L. for the motivation of this work.
\vspace{-0.1cm}
\ifCLASSOPTIONcaptionsoff
  \newpage
\fi


%
%
%
%

\bibliography{biblio}
\bibliographystyle{IEEEtran}

%




\end{document}